\documentclass[12pt]{article}
\usepackage{amsmath,amsthm,latexsym,amssymb,amsfonts,epsfig}

\makeatletter
\@addtoreset{equation}{section}
\makeatother

\newtheorem{Theorem}{Theorem}[section]

\def\be{\begin{equation}}
\def\ee{\end{equation}}
\def\ba{\begin{eqnarray}}
\def\ea{\end{eqnarray}}
\def\Nl{{\mathchoice
{\setbox0=\hbox{$\displaystyle\rm N$}\hbox{\hbox to0pt
{\kern0.4\wd0\vrule height0.9\ht0\hss}\box0}}
{\setbox0=\hbox{$\textstyle\rm N$}\hbox{\hbox to0pt
{\kern0.4\wd0\vrule height0.9\ht0\hss}\box0}}
{\setbox0=\hbox{$\scriptstyle\rm N$}\hbox{\hbox to0pt
{\kern0.4\wd0\vrule height0.9\ht0\hss}\box0}}
{\setbox0=\hbox{$\scriptscriptstyle\rm N$}\hbox{\hbox to0pt
{\kern0.4\wd0\vrule height0.9\ht0\hss}\box0}}}}
\def\Zl{{\mathchoice
{\setbox0=\hbox{$\displaystyle\rm Z$}\hbox{\hbox to0pt
{\kern0.4\wd0\vrule height0.9\ht0\hss}\box0}}
{\setbox0=\hbox{$\textstyle\rm Z$}\hbox{\hbox to0pt
{\kern0.4\wd0\vrule height0.9\ht0\hss}\box0}}
{\setbox0=\hbox{$\scriptstyle\rm Z$}\hbox{\hbox to0pt
{\kern0.4\wd0\vrule height0.9\ht0\hss}\box0}}
{\setbox0=\hbox{$\scriptscriptstyle\rm Z$}\hbox{\hbox to0pt
{\kern0.4\wd0\vrule height0.9\ht0\hss}\box0}}}}
\def\Ql{{\mathchoice
{\setbox0=\hbox{$\displaystyle\rm Q$}\hbox{\hbox to0pt
{\kern0.4\wd0\vrule height0.9\ht0\hss}\box0}}
{\setbox0=\hbox{$\textstyle\rm Q$}\hbox{\hbox to0pt
{\kern0.4\wd0\vrule height0.9\ht0\hss}\box0}}
{\setbox0=\hbox{$\scriptstyle\rm Q$}\hbox{\hbox to0pt
{\kern0.4\wd0\vrule height0.9\ht0\hss}\box0}}
{\setbox0=\hbox{$\scriptscriptstyle\rm Q$}\hbox{\hbox to0pt
{\kern0.4\wd0\vrule height0.9\ht0\hss}\box0}}}}
\def\Rl{{\mathchoice
{\setbox0=\hbox{$\displaystyle\rm R$}\hbox{\hbox to0pt
{\kern0.4\wd0\vrule height0.9\ht0\hss}\box0}}
{\setbox0=\hbox{$\textstyle\rm R$}\hbox{\hbox to0pt
{\kern0.4\wd0\vrule height0.9\ht0\hss}\box0}}
{\setbox0=\hbox{$\scriptstyle\rm R$}\hbox{\hbox to0pt
{\kern0.4\wd0\vrule height0.9\ht0\hss}\box0}}
{\setbox0=\hbox{$\scriptscriptstyle\rm R$}\hbox{\hbox to0pt
{\kern0.4\wd0\vrule height0.9\ht0\hss}\box0}}}}
\def\Cl{{\mathchoice
{\setbox0=\hbox{$\displaystyle\rm C$}\hbox{\hbox to0pt
{\kern0.4\wd0\vrule height0.9\ht0\hss}\box0}}
{\setbox0=\hbox{$\textstyle\rm C$}\hbox{\hbox to0pt
{\kern0.4\wd0\vrule height0.9\ht0\hss}\box0}}
{\setbox0=\hbox{$\scriptstyle\rm C$}\hbox{\hbox to0pt
{\kern0.4\wd0\vrule height0.9\ht0\hss}\box0}}
{\setbox0=\hbox{$\scriptscriptstyle\rm C$}\hbox{\hbox to0pt
{\kern0.4\wd0\vrule height0.9\ht0\hss}\box0}}}}
\def\Hl{{\mathchoice
{\setbox0=\hbox{$\displaystyle\rm H$}\hbox{\hbox to0pt
{\kern0.4\wd0\vrule height0.9\ht0\hss}\box0}}
{\setbox0=\hbox{$\textstyle\rm H$}\hbox{\hbox to0pt
{\kern0.4\wd0\vrule height0.9\ht0\hss}\box0}}
{\setbox0=\hbox{$\scriptstyle\rm H$}\hbox{\hbox to0pt
{\kern0.4\wd0\vrule height0.9\ht0\hss}\box0}}
{\setbox0=\hbox{$\scriptscriptstyle\rm H$}\hbox{\hbox to0pt
{\kern0.4\wd0\vrule height0.9\ht0\hss}\box0}}}}
\def\Ol{{\mathchoice
{\setbox0=\hbox{$\displaystyle\rm O$}\hbox{\hbox to0pt
{\kern0.4\wd0\vrule height0.9\ht0\hss}\box0}}
{\setbox0=\hbox{$\textstyle\rm O$}\hbox{\hbox to0pt
{\kern0.4\wd0\vrule height0.9\ht0\hss}\box0}}
{\setbox0=\hbox{$\scriptstyle\rm O$}\hbox{\hbox to0pt
{\kern0.4\wd0\vrule height0.9\ht0\hss}\box0}}
{\setbox0=\hbox{$\scriptscriptstyle\rm O$}\hbox{\hbox to0pt
{\kern0.4\wd0\vrule height0.9\ht0\hss}\box0}}}}

\title{Reduced Phase Space Quantization and Dirac Observables}
\author{T. Thiemann\thanks{thiemann@aei-potsdam.mpg.de, 
tthiemann@perimeterinstitute.ca} \\ \\
       MPI f. Gravitationsphysik, Albert-Einstein-Institut, \\
           Am M\"uhlenberg 1, 14476 Potsdam, Germany\\
                   \\  and \\ \\
Perimeter Institute f. Theoretical Physics,\\ 
Waterloo, ON N2L 2Y5, Canada}
\date{}
\begin{document}

\maketitle

\begin{abstract}
In her recent work, Dittrich generalized Rovelli's idea of partial 
observables to construct Dirac observables for constrained systems  
to the general case of an arbitrary 
first class constraint algebra with structure functions rather than 
structure constants. Here we use this framework and propose 
how to implement explicitly a reduced phase space quantization of a given 
system, at least in principle, without the need to compute the gauge 
equivalence classes. The degree of practicality of this programme depends 
on the choice of the partial observables involved. The (multi-fingered)
time evolution was shown to correspond to an automorphism on the set of 
Dirac observables so generated and interesting representations of the 
latter will be those for which a suitable preferred subgroup 
is realized unitarily. We sketch how such a programme might look like for 
General Relativity. 

We also observe that the ideas by Dittrich can be 
used in order to generate constraints equivalent to those of 
the Hamiltonian constraints for General Relativity such that they are 
spatially diffeomorphism invariant. This has the important consequence 
that one can now quantize the new Hamiltonian constraints on the 
partially reduced Hilbert space of spatially diffeomorphism invariant 
states, just as for the recently proposed Master constraint programme.
\end{abstract}

\section{Introduction}
\label{s1}

It is often stated that there are no Dirac observables known for General 
Relativity, except for the ten Poincar\'e charges at spatial infinity in 
situations with asymptotically flat boundary conditions. This is 
inconvenient for any quantization scheme because it is only the gauge 
invariant quantities, that is, the functions on phase space which have
weakly\footnote{We say that a relation holds weakly if it is an identity
on the constraint surface of the phase space where the constraints are 
satisfied.} 
vanishing Poisson brackets with the constraints, which have 
physical meaning and can be measured. These are precisely the (weak)
Dirac observables of the canonical formalism. The Dirac observables also
play a prominent role for the quantization at the technical level, because 
the ultimate physical Hilbert space must carry a representation of 
their Poisson algebra, no matter whether one follows a Dirac quantization 
scheme (reducing after quantizing) or a reduced phase space approach
(quantizing after reducing).

For General Relativity the identification of a suitably complete set of 
Dirac observables (that is, a set which encodes all the gauge invariant 
information about the system) is especially hard because the constraint 
algebra is not a Lie algebra: While it is a first class system, it does 
not close with structure constants but rather with structure functions,
that is, non trivial functions on phase space. This fact has obstructed 
the development of a representation theory of GR's constraint algebra
and hence the associated invariants. There are even obstruction theorems 
available in the literature \cite{0} which state the non existence of 
local Dirac observables (depending on a finite number of spatial 
derivatives) for GR. 

In \cite{5} a proposal for how to overcome the problem of structure 
functions for GR for the quantum theory in the context of Loop Quantum 
Gravity \cite{5a} was developed. The idea is to replace the algebra 
by a simpler but equivalent one which closes without structure functions.
The proposal was successfully applied to many test models of varying 
degree of complexity \cite{6}. Besides this quantum application also 
a meachnism to generate strong Dirac observables by the method of 
ergodic averaging was given. This avoids the obstruction theorem mentioned 
above because indeed the resulting observables do depend on an infinite 
number of spatial derivatives. However, strong Dirac observables are not 
particularly interesting for systems with structure functions, simply 
because there are probably not very many of them: If $\{C_j,O\}=0$ 
everywhere on phase space for all 
constraints $C_j$ and $\{C_j,C_k\}=f_{jk}\;^l C_l$ then by the Jacobi
identity $\{\{C_j,C_k\},O\}=C_l\{f_{jk}\;^l,O\}=0$ identically which due to 
the algebraic independence of the $C_j$ means that strong 
Dirac Observables also must satisfy the additional equations 
$\{f_{jk}\;^l,O\}=0$ and iterating like this it is quite possible that 
only the constants survive. 

In \cite{1} Dittrich addressed the problem of constructing weak Dirac 
observables for first class systems, possibly with structure constants.
Her construction is based on the notion of partial observables which to a 
large part is due to Rovelli \cite{2}. The idea is to use a relational 
point of view, namely to construct observables $F^\tau_{A,B}$ of the type: 
What is the 
value of a non -- ivariant function $A$ when under the gauge flow the non 
-- invariant function $B$ has arrived at the value $\tau$? The functions 
$A,B$ are here the partial observables and $F^\tau_{A,B}$ is called a 
complete observable. In \cite{2} it was shown that the complete observable 
is a strong Dirac observable for the case of a single constraint. 
The reason is that for a {\it single} constraint there is always an 
invariant combination between {\it two} functions: If we study the orbits
$\beta \mapsto A(\beta), B(\beta)$ of $A,B$ under the gauge flow 
parameterized by a Lagrange multiplier $\beta$ 
then we may use the value of $B$ as the parameter, i.e. we may invert the 
equation $B(\beta)=\tau$ for $\beta$ and insert that value into 
$A(\beta)$. The result is an invariant. In 
\cite{1} Dittrich generalized the idea to an arbitrary number $N$ 
of constraints by observing that similarly there is always a weakly
invariant combination between $N+1$ functions. This time we have to invert 
the $N-$parameter gauge flow for $N$ partial observables. 

In \cite{1} not only we find the proof that the result is always a
weak invariant but moreover an {\it explicit expression} is obtained
in terms of a formal power series. To the best of our knowledge, this  
is the first explicit expression and concrete algorithm for how to 
construct Dirac observables which moreover have a concrete (relational)
physical interpretation\footnote{There seems to be some overlap with 
\cite{1a}, however, the proofs are missing in that paper.}. 
On top of that, it is possible within this framework to calculate the
Poisson algebra of these Dirac observables and to implement a notion of 
multi -- fingered time \cite{8}, along the lines of Rovelli's ``evolving 
constants'', as certain Poisson automorphisms on that algebra.
While these findings were not derived like that, in retrospect the 
results of \cite{1} technically rest on the fact that, just like 
in \cite{5}, it is always possible to replace the constraint algebra by 
a simpler one.\\
\\
The present paper serves three purposes:\\ 
1.\\
The proofs of \cite{1} are very 
elegant and often use the definition of the complete 
obsservables in terms of the flow. In section \ref{s2} we review the  
parts of \cite{1} relevant for our purposes and give sometimes alternative 
proofs by directly working with the local,
explicit expressions and brutally working out the Poisson brackets.\\
2.\\
In section three we remark on the implications of \cite{1} for the 
purposes of the quantum 
theory. In particular, we sketch how to perform a reduced phase space 
quantization of the algebra of Dirac observables by making use of a 
suitable choice of partial observables. Such a choice is always available 
and we show that then, quite surprisingly, representations of a 
sufficiently large subalgebra of elementary Dirac observables are easily 
available.
The multi -- fingered time evolution can be implemented unitarily provided 
one can quantize the corresponding Hamiltonian generators as aself -- 
adjoint operators in those representations. While it is possible to 
identify those Hamiltonians, their explicit form in terms of elementary 
Dirac observables will be very complicated in general. 

In section \ref{s4} we also sketch how such a quantization scheme might 
look concretely when 
applied to General Relativity coupled to the Standard model and in 
section \ref{s5} we comment on the important physical differences between 
the 
reduced phase space quantization and the Dirac contraint quantization
that is currently performed in LQG.\\
3.\\
In section \ref{s6} we combine the ideas of \cite{1} with those of 
\cite{5} by showing how the Master Constraint Programme for 
General Relativity can be used in order to provide {\it spatially 
diffeomorphism invariant Hamiltonian constraints}. The important 
consequence of this is that in the constraint quantization one can 
implement the new Hamiltonian constraints on the spatially diffeomorphism 
invariant Hilbert space which is not possible for the old constraints 
because for those the spatial diffeomorphism subalgebra is not an ideal. 
As a consequence, the algebra of the new Hamiltonian constraints on 
the spatially dfiffeomorphism invariant Hilbert space then closes on 
itself (albeit with structure functions rather than structure constants
in general). This might pose an attractive alternative to the 
previous Hamiltonian constraint quantization \cite{7}. \\
\\
We conclude in section \ref{s7}.

\section{Review of the Classical Framework}
\label{s2}

We summarize here the work of \cite{1} and sometimes give alternative 
proofs. These use only local considerations, hence no global 
assumptions are made. On the other hand, the results are only locally 
valid (e.g. the clock variables must provide a good coordinatization of 
the gauge orbits). This is enough from a physical perspective since 
physical clocks are not expected to be good coordinates everywhere on 
phase space.

\subsection{Partial and Weak Dirac Observables}
\label{s2.1}

Let $C_j,\;j\in {\cal I}$ be a system of first class constraints on a 
phase space $\cal M$ with (strong) symplectic structure given by a 
Poisson bracket $\{.,.\}$ where the index set has countable cardinality. 
This 
includes the case of a field theory for which the constraints are usually
given in the local form $C_\mu(x),\; x\in \sigma,\;\mu=1,..,n<\infty$ 
where $\sigma$ is a spatial, $D-$dimensional manifold corresponding to the 
initial value 
formulation and $\mu$ are some tensorial and/or Lie algebra indices. 
This can be seen by choosing a basis $b_I$ of the Hilbert space 
$L_2(\sigma,d^Dx)$ consisting of smooth functions of compact support 
and defining $C_j:=\int_\sigma d^Dx b_I(x) C_\mu(x)$ with 
$j:=(\mu,I)$. We assume the most general situation, namely that 
$\{C_j,C_k\}=f_{jk}\;^l C_l$ closes with structure functions, that is,
$f_{jk}\;^l$ can be non -- trivial functions on $\cal M$.

The partial observable Ansatz to generate Dirac observables is now as 
follows:
%\footnote{The idea is due to Rovelli \cite{2} but to the best of 
%our knowledge has been stated only for the case of finitely many 
%Abelean constraints. The extension to general case is due to Dittrich
%\cite{2}.}
Take as many functions on phase space $T_j,\;j\in {\cal I}$ as 
there are constraints. These functions have the purpose of providing a 
local (in phase space) coordinatization of the gauge orbit 
$[m]$ of any point $m$ in phase space, 
at least in a neighbourhood of the constraint surface
$\overline{{\cal M}}=\{m\in {\cal M};\;C_j(m)=0\;\forall j\in {\cal I}\}$.
The gauge orbit $[m]$ of $m$ is given by 
$[m]:=\{\alpha_{\beta_1}\circ..\circ\alpha_{\beta_N}(m);\;
N<\infty,\;\beta^j_k\in \Rl,\;k=1,..,N,\;j\in {\cal I}\}$. 
Here $\alpha_\beta$ is the canonical transformation (automorphism of 
$(C^\infty({\cal M}),\;\{.,.\})$ generated by the Hamiltonian vector field 
$\chi_\beta$ of $C_\beta:=\beta^j C_j$, that is 
$\alpha_\beta(f):=\exp(\chi_\beta)\cdot f$. (Notice that 
if the system would have structure constants, then it would be sufficient 
to choose $N=1$.) 

In other words, we assume that it is possible to find 
functions $T_j$ such that each $m\in \overline{{\cal M}}$ 
is completely specified by $[m]$ and by the $T_j(m)$. This means that if 
the value $\tau_j$ is in the range of $T_j$ then the gauge fixing surface 
$\overline{{\cal M}}_\tau:=\{m\in\overline{{\cal M}};\;T_j(m)=\tau_j\}$
intersects each $m$ in precisely one point. In practice this is usually
hard to achieve globally on $\overline{{\cal M}}$ due to the possibility 
of Gribov copies but here we are only interested in local considerations. 
It follows that the matrix $A_{jk}:=\{C_j,T_k\}$ must be locally 
invertible so that the condition $[\alpha_\beta(T_j)](m)
=T_j(\alpha_\beta(m))=\tau_j$
can be inverted for $\beta$ (given $m'\in m$ we may write it in the 
form $[\alpha_\beta(m)]_{|\beta=B(m)}$ for some $B(m)$ which may depend 
on $m$).    

Take now another function $f$ on phase space. Then the weak Dirac 
observable
$F^\tau_{f,T}$ associated to the partial observables 
$f,T_j,\;j\in {\cal I}$ is defined by 
\be \label{2.1}
(F^\tau_{f,T})(m):=[f(\alpha_\beta(m))]_{|\beta=B^\tau_T(m)},\;\;
[T_j(\alpha_\beta(m))]_{|\beta=B^\tau_T(m)}=\tau_j
\ee
The physical interpretation of $F^\tau_{f,T}$ is that it is the value 
of $f$ at those ``times'' $\beta_j$ when the ``clocks'' $T_j$ take the 
values $\tau_j$.

In \cite{1} a proof was given that (\ref{2.1}) is indeed invariant under 
the flow automorphisms $\alpha_\beta$ despite the fact that the 
$\alpha_\beta$ do not form a group of automorphisms in the case of 
structure functions. This is quite astonishing given 
the fact that the direct proof for the case of a single constraint can 
be easily repeated only in the case that the constraints are mutually 
commuting. Then an explicit expression was derived using the system of 
partial differential equations (in the parameters $\tau_j$) that the 
$F^\tau_{f,T}$ satisfy. 

We will now derive that same explicit expression 
from an Ansatz for a Taylor expansion. Namely, on the gauge cut 
$\overline{{\cal M}}_\tau$ the function $F^\tau_{f,T}$ equals $f$
since then $B^\tau_T(m)=0$. Away from this section, $F^\tau_{f,T}$
can be expanded into a Taylor series\footnote{In other words, 
$F^\tau_{f,T}$
is the gauge invariant extension of the restriction of $f$ to 
$\overline{{\cal M}}_\tau$ mentioned in \cite{3} for which however no 
explicit expression was given there.}. Thus we make the Ansatz
\be \label{2.2}
F^\tau_{f,T}=\sum_{\{k_j\}_{j\in {\cal I}}=0}^\infty\;
\prod_{j\in {\cal I}} \frac{(\tau_j-T_j)^{k_j}}{k_j !}
f_{\{k_j\}_{j\in {\cal I}}}
\ee
with $f_{\{k_j\}=\{0\}}=f$. We assume that (\ref{2.2}) converges 
absolutely on a open set $S$ and is continuous there, hence is
uniformly bounded on $S$. We may then interchange summation and 
differentiation on $S$ and compute
\ba \label{2.3}
\{C_l,F^\tau_{f,T}\} &=& \sum_{\{k_j\}_{j\in {\cal I}}=0}^\infty\;
\prod_{j\in {\cal I}} \frac{(\tau_j-T_j)^{k_j}}{k_j !}
\times\nonumber\\
&& +
[\sum_{m\in {\cal I}} -A_{l,m} f_{\{k'_j(m)\}_{j\in {\cal I}}}
+\{C_l,f_{\{k_j\}_{j\in {\cal I}}}]
\ea
where $k'_j(m)=k_j$ for $j\not=m$ and $k'_m(m)=k_m+1$. Setting (\ref{2.3})
(weakly) to zero leads to a recursion relation with the formal solution
\be \label{2.4}
f_{\{k_j\}_{j\in {\cal I}}}=\prod_{j\in {\cal I}} (X'_j)^k_j \cdot f,\;
X'_j\cdot f=\sum_{k\in {\cal I}} (A^{-1})_{jk} \{C_k,f\}
\ee
Expression (\ref{2.4}) is formal because we did not specify the order of 
application of the vector fields $X'_j$. We will now show that, as a weak
identity, the order in (\ref{2.4}) is irrelevant. To see this, let us 
introduce the equivalent constraints (at least on $S$)
\be \label{2.5}
C'_j:=\sum_{k\in {\cal I}} (A^{-1})_{jk} C_k
\ee
and notice that with the Hamiltonian vector fields $X_j\cdot f=\{C'_j,f\}$
we have $X'_{j_1}..X'_{j_n} \cdot f\approx X_{j_1}..X_{j_n} \cdot f$ for 
any $j_1,..j_n$
due to the first class property of the constraints. Here and what follows 
we write $\approx$ for a relation that becomes an identity on 
$\overline{{\cal M}}$. Then we can make the following surprising 
observation.
\begin{Theorem} \label{th2.1} ~\\
Let $C_j$ be a system of first class constraints and $T_j$ be any 
functions such that the matrix $A$ with entries 
$A_{jk}:=\{C_j,T_k\}$ is invertible on some open set $S$ intersecting 
the constraint surface. Define the 
equivalent $C'_j$ constraints (\ref{2.5}). Then their Hamiltonian vector
fields $X_j:=\chi_{C'_j}$ are mutually weakly commuting.
\end{Theorem}
Proof of theorem \ref{th2.1}:\\
The proof consists of a straightforward computation and exploits the 
Jacobi identity. Abbreviating $B_{jk}:=(A^{-1})_{jk}$ we have
\ba \label{2.6}
&& \{C'_j,\{C'_k,f\}\}-\{C'_k,\{C'_j,f\}\}
\approx \sum_{m,n} B_{jm}\{C_m, [B_{kn} \{C_n,f\}+C_n \{B_{kn},f\}]\}
- j \leftrightarrow k 
\nonumber\\
&\approx& \sum_{m,n} B_{jm}[\{C_m,B_{kn}\} \{C_n,f\}+B_{kn} 
\{C_m,\{C_n,f\}\}]
- j \leftrightarrow k 
\nonumber\\
&=&\sum_{m,n} B_{jm}[-\sum_{l,i} B_{kl} \{C_m,A_{li}\} B_{in} \{C_n,f\}
+B_{kn} \{C_m,\{C_n,f\}\}]
- j \leftrightarrow k 
\nonumber\\
&=&\sum_{m,n} B_{jm}[-\sum_{l,i} B_{kl} B_{in} \{C_n,f\} 
(\{C_m,\{C_l,T_i\}\}-\{C_l,\{C_m,T_i\}\}) 
+B_{kn} (\{C_m,\{C_n,f\}\}-\{C_n,\{C_m,f\}\})]
\nonumber\\
&=&\sum_{m,n} B_{jm}[\sum_{l,i} B_{kl} B_{in} \{C_n,f\} 
\{T_i,\{C_m,C_l\}\} -B_{kn} (\{f,\{C_m,C_n\}\}]
\nonumber\\
&\approx &\sum_{m,n} B_{jm}[-\sum_{l,i,p} B_{kl} B_{in} \{C_n,f\} 
f_{ml}\;^p A_{pi}+B_{kn}\sum_l f_{mn}\;^l \{C_l,f\}]
\nonumber\\
&= &\sum_{m,n,l} B_{jm}[-B_{kl} \{C_n,f\} 
f_{ml}\;^n +B_{kn} f_{mn}\;^l \{C_l,f\}]
\nonumber\\
&=& 0
\ea
Due to 	
\be \label{2.7}
\{C'_j,\{C'_k,f\}\}-\{C'_k,\{C'_j,f\}\}
=\{\{C'_j,C'_k\},f\}\approx f'_{jk}\;^l \{C'_l,f\} \approx 0
\ee
this means that the structure functions $f'_{jk}\;^l$ with respect to the 
$C'_j$ are weakly vanishing, that is, themselves proprotional to the 
constraints. \\
$\Box$\\
We may therefore write the Dirac observable generated by $f,T_j$ indeed as 
\be \label{2.8}
F^\tau_{f,T}=\sum_{\{k_j\}_{j\in {\cal I}}=0}^\infty\;
\prod_{j\in {\cal I}} \frac{(\tau_j-T_j)^{k_j}}{k_j !}\;
\prod_{j\in {\cal I}} (X_j)^{k_j}\cdot f
\ee
Expression (\ref{2.8}) is, despite the obvious convergence issues to be 
checked in the concrete application, remarkably simple. Of course, 
especially in field theory it will not be possible to calculate it exactly
and already the computation of the inverse $A^{-1}$ may be hard, depending 
on the choice of the $T_j$. However, for points close to the gauge cut 
expression (\ref{2.8}) is rapidly converging and one may be able to 
do approximate calculations.\\
\\
Remark:\\
Let $\alpha'_\beta(f):=\exp(\sum_j \beta_j X_j)\cdot f$ be the gauge flow 
generated by the new constraints $C'_j$ for real valued gauge parameters 
$\beta_j$. We easily calculate $\alpha'_\beta(T_j)\approx T_j+\beta_j$.
The condition $\alpha'_\beta(T_j)=\tau_j$ can therefore be easily inverted 
to $\beta_j\approx \tau_j-T_j$. Hence the complete observable prescription
with respect to the new constraints $C'_j$
\be \label{2.8a}
F^\tau_{f,T}:=[\alpha'_\beta(f)]_{|\alpha'_\beta(T)=\tau}
\ee
weakly coincides with (\ref{2.8}).

\subsection{Poisson algebra of Dirac Observables}
\label{s2.2}

In \cite{3} we find the statement that the Poisson brackets among the 
Dirac observables obtained as the gauge invariant extension off 
$\overline{{\cal M}}_\tau$ of the respective restrictions to the gauge cut 
of functions $f,g$ is weakly given by the gauge invariant extension of 
their Dirac bracket with respect to the associated gauge fixing functions.
However, a proof of that statement could nowhere be found by the present 
author. Expression (\ref{2.8}) now enables us to give an explicit, local 
proof (modulo convergence issues). See \cite{1} for an alternative one.
\begin{Theorem} \label{th2.2} ~\\
Let $F^\tau_{f,T}$ be defined as in (\ref{2.8}) with respect to partial 
observables $T_j$. Introduce the gauge 
conditions $G_j:=T_j-\tau_j$ and consider the system of second class 
constraints $C_{1j}:=C_j,\;C_{2j}:=G_j$ and abbreviate 
$\mu=(I,j),\;I=1,2$. Introduce the Dirac bracket
\be \label{2.9}
\{f,f'\}^\ast:=\{f,f'\}-\{f,C_\mu\} K^{\mu\nu} \{C_\nu,f'\}
\ee
where 
$K_{\mu\nu}=\{C_\mu,C_\nu\},\;K^{\mu\rho}K_{\rho\nu}=\delta^\mu_\nu$.
Then 
\be \label{2.10}
\{F^\tau_{f,T},F^\tau_{f',T}\}\approx F^\tau_{\{f,f'\}^\ast,T}
\ee
\end{Theorem}
Proof of theorem \ref{th2.2}:\\
Let us introduce the abbreviations 
\be \label{2.11}
Y_{\{k\}}=\prod_j \frac{(\tau_j-T_j)^{k_j}}{k_j !},\;\;
f_{\{k\}}=\prod_j (X_j)^{k_j}\cdot f,\;\;
\sum_{\{k\}}=\sum_{k_1,k_2,..=0}^\infty
\ee
We have 
\ba \label{2.12}
&& \{F^\tau_{f,T},F^\tau_{f',T}\} 
=
\sum_{\{k\},\{l\}} \;\{Y_{\{k\}} f_{\{k\}},Y_{\{l\}} f'_{\{l\}}\}
\nonumber\\
&\approx&
\sum_{\{k\},\{l\}} \;Y_{\{k\}}\;Y_{\{l\}} 
[\{ f_{\{k\}}, f'_{\{l\}}\}
-\sum_j (X_j\cdot f)_{\{k\}} \{T_j,f'_{\{l\}}\}
\nonumber\\
&& +\sum_j (X_j\cdot f')_{\{l\}} \{T_j,f_{\{k\}}\}
+\sum_{j,m} (X_j\cdot f)_{\{k\}}(X_m\cdot f')_{\{l\}} \{T_j,T_m\}]
\nonumber\\
&=&
\sum_{\{n\}} \;Y_{\{n\}} \sum_{\{k\};\;k_l\le n_l} 
\prod_l \; \left( \begin{array}{c} n_l \\ k_l \end{array} \right) \;  
[\{ f_{\{k\}}, f'_{\{n-k\}}\}
-\sum_j (X_j\cdot f)_{\{k\}} \{T_j,f'_{\{n-k\}}\}
\nonumber\\
&& +\sum_j (X_j\cdot f')_{\{n-k\}} \{T_j,f_{\{k\}}\}
+\sum_{j,m} (X_j\cdot f)_{\{k\}} (X_m\cdot f')_{\{n-k\}} \{T_j,T_m\}]
\ea
By definition of a Hamiltonian vector field we have 
$X_j\{f,f'\}=\{X_j f,f'\}+\{f,X_j f'\}$. Thus, by the (multi) Leibniz rule 
\be \label{2.13}
\prod_l (X_l)^n_l \{f,f'\}=   
\sum_{\{k\};\;k_l\le n_l} 
\prod_l \; \left( \begin{array}{c} n_l \\ k_l \end{array} \right) \;  
[\{ f_{\{k\}}, f'_{\{n-k\}}\}
\ee
is already the first term we need. It therefore remains to show that 
\ba \label{2.14}
&& \prod_l (X_l)^n_l [\{f,f'\}^\ast-\{f,f'\}]\approx   
\sum_{\{k\};\;k_l\le n_l} 
\prod_l \; \left( \begin{array}{c} n_l \\ k_l \end{array} \right) \;  
[-\sum_j (X_j\cdot f)_{\{k\}} \{T_j,f'_{\{n-k\}}\}
\nonumber\\
&& +\sum_j (X_j\cdot f')_{\{n-k\}} \{T_j,f_{\{k\}}\}
+\sum_{j,m} (X_j\cdot f)_{\{k\}} (X_m\cdot f')_{\{n-k\}} \{T_j,T_m\}]
\ea
We will do this by multi induction over $N:=\sum_l n_l$.\\
The case $N=0$ reduces to the claim
\ba \label{2.15}
\{f,f'\}^\ast-\{f,f'\}
&&  \approx   
-\sum_j (X_j\cdot f) \{T_j,f'\}
\nonumber\\
&& +\sum_j (X_j\cdot f') \{T_j,f\}
+\sum_{j,m} (X_j\cdot f) (X_m\cdot f') \{T_j,T_m\}
\ea
To compute the Dirac bracket explicitly we must invert the matrix
$K_{Jj,Kk}$ with entries $K_{1j,1k}=\{C_j,C_k\}=f_{jk}\;^l C_l\approx 0$,
$K_{1j,2k}=\{C_j,T_k\}=A_{jk}=-K_{2k,1j}$ and $K_{2j,2k}=\{T_j,T_k\}$.
By definition $\sum_{L,l} K^{Jj,Ll} K_{Ll,Kk}=\delta^J_K\delta^j_k$
therefore $K^{1j,1k}\approx\sum_{m,n} (A^{-1})_{mj} \{T_m,T_n\} 
(A^{-1})_{nk}$,
$K^{1j,2k}\approx -(A^{-1})_{kj}\approx -K^{2k,1j}$ and 
$K^{2j,2k}\approx 0$. It follows 
\ba \label{2.16}
&&-\{f,f'\}^\ast+\{f,f'\}
=\{f,C_j\} K^{1j,1k} \{C_k,f'\}
+\{f,C_j\} K^{1j,2k} \{T_k,f'\}
\nonumber\\
&& +\{f,T_j\} K^{2j,1k} \{C_k,f'\}
+\{f,T_j\} K^{2j,2k} \{T_k,f'\}
\nonumber\\
&\approx &
\sum_{m,n} \{f,C_j\} (A^{-1})_{mj} \{T_m,T_n\} 
(A^{-1})_{nk} \{C_k,f'\}
-\{f,C_j\} (A^{-1})_{kj} \{T_k,f'\}
+\{f,T_j\} (A^{-1})_{jk}\{C_k,f'\}
\nonumber\\
&\approx &
-\sum_{m,n} (X_m\cdot f)\{T_m,T_n\} (X_n\cdot f')
+ (X_k\cdot f) \{T_k,f'\}
- (X_k \cdot f') \{T_k,f\} 
\ea
which is precisely the negative of (\ref{2.15}).\\
\\
Suppose then that we have proved the claim for every configuration 
$\{n_l\}$ such that $\sum_l n_l\le N$. Any configuration with $N+1$ arises 
from a configuration with $N$ by raising one of the $n_l$ by one unit, say
$n_j\to n_j+1$. Then, by assumption 
\ba \label{2.17}
&& X_j \prod_l (X_l)^n_l [\{f,f'\}^\ast-\{f,f'\}]
\nonumber\\
&\approx &  
X_j \sum_{\{k\};\;k_l\le n_l} 
\prod_l \; \left( \begin{array}{c} n_l \\ k_l \end{array} \right) \;  
[-\sum_l (X_l\cdot f)_{\{k\}} \{T_l,f'_{\{n-k\}}\}
\nonumber\\
&& +\sum_l (X_l\cdot f')_{\{n-k\}} \{T_l,f_{\{k\}}\}
+\sum_{l,m} (X_l\cdot f)_{\{k\}} (X_m\cdot f')_{\{n-k\}} \{T_l,T_m\}]
\nonumber\\
&\approx&
\sum_{\{k\};\;k_l\le n_l} 
\prod_l \; \left( \begin{array}{c} n_l \\ k_l \end{array} \right) \;  
\times\nonumber\\
&&
-\sum_l [
(X_l\cdot f)_{\{k^j\}} \{T_l,f'_{\{n-k\}}\}
+(X_l\cdot f)_{\{k\}} \{T_l,f'_{\{n^j-k\}}\}
+(X_l\cdot f)_{\{k\}} \{X_j\cdot T_l,f'_{\{n-k\}}\}]
\nonumber\\
&& +\sum_l [
(X_l\cdot f')_{\{k^j\}} \{T_l,f_{\{n-k\}}\}
+(X_l\cdot f')_{\{k\}} \{T_l,f_{\{n^j-k\}}\}
+(X_l\cdot f')_{\{k\}} \{X_j\cdot T_l,f_{\{n-k\}}\}]
\nonumber\\
&& 
+\sum_{l,m} [
(X_l\cdot f)_{\{k^j\}} (X_m\cdot f')_{\{n-k\}} \{T_l,T_m\}
+(X_l\cdot f)_{\{k\}} (X_m\cdot f')_{\{n^j-k\}} \{T_l,T_m\}
\nonumber\\
&& +(X_l\cdot f)_{\{k\}} (X_m\cdot f')_{\{n-k\}} 
(\{X_j T_l,T_m\}+\{T_l,X_j T_m\})]]
\ea
where $\{k^j\}$ coincides with $\{k\}$ except that $k_j\to k_j+1$ and 
similar for $\{n^j\}$. By the multi binomial theorem the first two terms 
in each of the three sums in the last equality combine precisely to what 
we need. Hence 
it remains to show that 
\ba \label{2.18}
0 &\approx & \sum_{\{k\};\;k_l\le n_l} 
\prod_l \; \left( \begin{array}{c} n_l \\ k_l \end{array} \right) \;  
\times\nonumber\\
&&
[-\sum_l (X_l\cdot f)_{\{k\}} \{X_j\cdot T_l,f'_{\{n-k\}}\}
+\sum_l (X_l\cdot f')_{\{k\}} \{X_j\cdot T_l,f_{\{n-k\}}\}
\nonumber\\
&& 
+\sum_{l,m} 
(X_l\cdot f)_{\{k\}} (X_m\cdot f')_{\{n-k\}} 
(\{X_j T_l,T_m\}+\{T_l,X_j T_m\})]
\ea
We have 
\be \label{2.19}
X_j \cdot T_l=\delta_{jl}+\sum_m C_m \{(A^{-1})_{jm},T_l\}]
=:\delta_{jl}+\sum_m C_m B_{jlm}
\ee
Hence
\be \label{2.20}
\{X_j\cdot T_l,g\}\approx \sum_{m,n} B_{jlm} A_{mn} (X_n\cdot g)
=:\sum_n D_{jln} (X_n \cdot g)
\ee
Next, using (\ref{2.19}) and (\ref{2.20})
\be \label{2.21}
\{X_j T_l,T_m\}+\{T_l,X_j T_m\}
\approx \sum_n (B_{jln} A_{nm}-B_{jmn} A_{nl})=D_{jlm}-D_{jml}
\ee
We now can simplify the right hand side of (\ref{2.18}) 
\ba \label{2.22}
&& \sum_{\{k\};\;k_l\le n_l} 
\prod_l \; \left( \begin{array}{c} n_l \\ k_l \end{array} \right) \;  
\times\nonumber\\
&& \sum_{l,m} D_{jlm}
[-(X_l\cdot f)_{\{k\}} (X_m \cdot f'_{\{n-k\}})
 +(X_l\cdot f')_{\{k\}} (X_m\cdot f_{\{n-k\}})
\nonumber\\
&& 
+[D_{jlm}-D_{jml}]
(X_l\cdot f)_{\{k\}} (X_m\cdot f')_{\{n-k\}}]
\nonumber\\
&& \sum_{l,m} D_{jlm} \prod_i (X_i)^{n_i} [
[-(X_l\cdot f) (X_m \cdot f')
 +(X_l\cdot f') (X_m\cdot f)
+(X_l\cdot f) (X_m\cdot f')
-(X_m\cdot f) (X_l\cdot f')]
\nonumber\\
&=& 0
\ea
as claimed. Notice that by using the Jacobi identity we also have 
$D_{jkl}=D_{jlk}$ so the two terms in the second and third 
line of (\ref{2.22}) even vanish separately (important 
for the case that $\{T_j,T_k\}=0$). \\
$\Box$\\
We can now rephrase theorem \ref{th2.2} as follows:\\
Consider the map
\be \label{2.23}
F^\tau_T:\;(C^\infty({\cal M}),\{.,.\}_T^\ast) \to 
(D^\infty({\cal M}),\{.,.\}_T^\ast);\;f\mapsto F^\tau_{f,T}
\ee
where $D^\infty({\cal M})$ denotes the set of smooth, weak Dirac 
observables and $\{.,.\}_T^\ast$ is the Dirac bracket with respect 
to the gauge fixing functions $T_j$. Then theorem \ref{th2.2} says that 
$F^\tau_T$ is a weak Poisson homomorphism (i.e. a homomorphism 
on the constraint surface). To see this, notice that for
(weak) Dirac observables the Dirac bracket coincides weakly with the 
ordinary Poisson bracket. Moreover, the map $F^\tau_T$ is linear and 
trivially
\ba \label{2.24}
&& F^\tau_{f,T} \;F^\tau_{f',T}
=\sum_{\{k\},\{l\}} Y_{\{k\}} Y_{\{l\}} f_{\{k\}} f'_{\{l\}}
\nonumber\\
&=& \sum_{\{n\}} Y_{\{n\}} 
\sum_{\{k\};\;k_l\le n_l} 
\prod_l \; \left( \begin{array}{c} n_l \\ k_l \end{array} \right) \;  
f_{\{k\}} f'_{\{n-k\}}
\nonumber\\
&\approx & \sum_{\{n\}} Y_{\{n\}} \prod_l (X_l)^{n_l} (f \;f')
=F^\tau_{f f',T}
\ea
(We can make the homomorphism exact by dividing both $C^\infty({\cal M})$
and $D^\infty({\cal M})$ by the ideal (under pointwise addition and 
multiplication) of smooth functions vanishing on the constraint surface,
see \cite{1}). We will use this important fact, to the best of our 
knowledge first observed in \cite{1}, for a new proposal towards 
quantization. Notice, that $F^\tau_T$ is onto because $F^\tau_{f,T}\approx 
f$ if $f$ is already a weak Dirac observable.

\subsection{Evolving Constants}
\label{s2.3}

The whole concept of partial observables was invented in order to remove 
the following conceptual puzzle:\\
In a time reparameterization invariant system such as General Relativity 
the formalism asks us to find the time reparameterization invariant
functions on phase space. However, then ``nothing happens'' in the theory,
there is no time evolution, in obvious contradiction to what we observe. 
This 
puzzle is removed by using the partial observables by taking the 
relational point of view: The partial observables $f,T_j$ can be measured 
but not predicted. However, we can predict $F^\tau_{f,T}$, it has the 
physical interpretation of giving the value of $f$ when the $T_j$ assume 
the values $\tau_j$. In constrained field theories we thus arrive at 
the multi fingered time picture, there is no preferred time but there 
are infinitely many. Accordingly, we define a multi -- fingered time 
evolution on the image of the maps $F^\tau_T$ by
\be \label{2.24a}
\alpha^\tau:\; F^{\tau^0}_T(C^\infty({\cal M}))\to 
F^{\tau+\tau^0}_T(C^\infty({\cal M}));\;
F^{\tau^0}_{f,T}\mapsto F^{\tau+\tau^0}_{f,T}
\ee
As defined, $\alpha^\tau$ forms an Abelean group. However, it has even 
more interesting properties:
\ba \label{2.25}
F^{\tau+\tau_0}_{f,T}
&=&
\sum_{\{n\}} \prod_j \frac{(\tau_j+\tau^0_j-T_j)^{n_j}}{n_j !}
\;\prod_j X_j^{n_j} \cdot f
\nonumber\\
&\approx&
\sum_{\{n\}} 
\sum_{\{k\};\;k_l\le n_l} 
\prod_l \; \frac{1}{n_l !} \left( \begin{array}{c} n_l \\ k_l \end{array} 
\right) \;  
\prod_j (\tau^0_j-T_j)^{k_j} \tau_j^{n_j-k_j}
\;\prod_j X_j^{k_j} \; X_j^{n_j-k_j} \cdot f
\nonumber\\
&\approx &
\sum_{\{k\}}
\prod_j \frac{(\tau^0_j-T_j)^{k_j}}{k_j!} 
\;\prod_j X_j^{k_j}\; \cdot [\sum_{\{l\}}
\frac{\tau_j^{l_j}}{l_j !}
\;\prod_j X_j^{l_j}] \cdot f
\nonumber\\
&=& F^{\tau_0}_{\alpha'_\tau(f),T}
\ea
where $\alpha'_\tau(f)$ is the automorphism on $C^\infty({\cal M})$ 
generated by the Hamiltonian vector field of 
$\sum_j \tau_j C'_j$ with the equivalent constraints $C'_j=\sum_k 
(A^{-1})_{jk} C_k$. This is due to the multi -- nomial theorem
\ba \label{2.26}
\alpha'_\tau(f)&=&
\sum_{n=0}^\infty \frac{1}{n!} (\sum_j \tau_j X_j)^n\cdot f
\nonumber\\
&=&
\sum_{n=0}^\infty \frac{1}{n!} \sum_{j_1,..,j_n} 
\prod_{k=1}^n \tau_{j_k} X_{j_k}\;\cdot f
\nonumber\\
&=&
\sum_{n=0}^\infty \frac{1}{n!} \sum_{\{k\};\sum_j k_j=n} 
\frac{n!}{\prod_j (k_j)!} \prod_j \tau_j^{k_j} \;\prod_j X_j^{k_j}\;\cdot 
f
\nonumber\\
&=&
\sum_{\{k\}} \prod_j \frac{\tau_j^{k_j}}{k_j!} \;\prod_j X_j^{k_j}\;\cdot 
f
\ea
Thus, our time evolution on the observables is induced by a gauge 
transformation on the partial observables.
From this observation it follows, together with the weak homomorphism
property, that 
\ba \label{2.27}
&& \{\alpha^\tau(F^{\tau_0}_{f,T}),\alpha^\tau(F^{\tau_0}_{f',T})\}
=\{F^{\tau_0+\tau}_{f,T},F^{\tau_0+\tau}_{f',T}\}
\nonumber\\
&\approx& F^{\tau_0+\tau}_{\{f,f'\}^\ast,T}=
\alpha^\tau(F^{\tau_0}_{\{f,f'\}^\ast,T}) 
\nonumber\\
&\approx& 
\alpha^\tau(\{F^{\tau_0}_{f,T},F^{\tau_0}_{f',T}\})
\ea
In other words, $\tau\mapsto \alpha^\tau$ is a weak, Abelean, multi -- 
parameter group of automorphisms on the image of each map 
$F^{\tau_0}_{f,T}$. This is in strong analogy to the properties of the 
one parameter group of automorphisms on phase space generated by a true 
Hamiltonian. Also this observation, in our opinion due to \cite{1}, will 
be used for a new proposal towards quantization.

\section{Reduced Phase Space Quantization of the Algebra of Dirac
Observables and Unitary Implementation of the Multi -- Fingered Time
Evolution}
\label{s3}

We will now describe our proposal. We assume that it is possible to
to choose the functions $T_j$ as canonical coordinates. In other words,
we choose a canonical coordinate system consisting of canonical pairs
$(q^a,p_a)$ and $(T_j,P^j)$ where the first system of coordinates has 
vanishing Poisson brackets with the second so that the only non vanishing
brackets are $\{p_a,q^b\}=\delta_a^b,\;\{P^j,T_k\}=\delta^j_k$. 
(In field theory the label set of the $a,b,..$ will be indefinite 
corresponding to certain smeared quantities of the canonical fields).
The 
virtue of this assumption is that the Dirac bracket reduces to the 
ordinary Poisson bracket on functions which depend only on $q^a,p_a$. 
We will shortly see why this is important. We define with 
$F_T:=F^0_T$ the weak Dirac observables at multi fingered time $\tau=0$
(or anny other fixed allowed value of $\tau$).
\be \label{3.1}
Q^a:=F_T(q^a),\; P_a:=F_T(p_a)
\ee
Notice that $F^\tau_{T_j,T}\approx \tau_j$, so the Dirac observable 
corresponding to $T_j$ is just a constant and thus not very interesting
(but evolves precisely as a clock). Likewise $F\tau_{C_j,T}\approx 0$
is not very interesting. Since at least locally we can solve the 
constraints $C_j$ for the momenta $P^j$, that is 
$P^j\approx E_j(q^a,p_a,T_k)$ and $F_T$ is a homomorphism with respect to
pointwise operations we have 
\be \label{3.2}
F_T(P_j)\approx 
E_j(F_T(q^a),F_T(p_a),F_T(T_k))\approx E_j(Q^a,P_a,\tau_k)
\ee
and thus also does not give rise to a Dirac observable which we could not 
already construct from $Q^a,P_a$. The importance of our assumption is now 
that due to the homomorphism property
\be \label{3.3}
\{P_a,Q^b\}\approx 
F^0_{\{p_a,q^b\}^\ast,T}=F^0_{\delta_a^b,T}=\delta_a^b,\;\;
\{Q^a,Q^b\}\approx\{P_a,P_b\}\approx 0
\ee
In other words, even though the functions $P_a,Q^a$ are very complicated 
expressions in terms of $q^a,p_a,T_j$ they have nevertheless canonical 
brackets at least on the constraint surface. If we would have had to use 
the Dirac bracket then this would not be the case and the algebra among 
the $Q^a,P_a$ would be too complicated and no hope would exist towards its 
quantization. However, under our assumption there is now a chance.

Now reduced phase space quantization consists in quantizing the subalgebra 
of $\cal D$, spanned by our preferred Dirac observables $Q^a, P_a$ 
evaluated on the constraint surface. 
As we have just seen, the algebra $\cal D$ itself is 
given by the Poisson algebra of the functions of the $Q^a,P_a$ evaluated 
on the constraint surface. Hence all the 
weak equalities that we have derived now become exact. We are therefore
looking for a representation $\pi:\;{\cal D} \to {\cal L}({\cal H})$
of that subalgebra of $\cal D$ as self -- adjoint, linear operators on a 
Hilbert space such 
that $[\pi(P_a),\pi(Q^b)]=i\hbar \delta_a^b$. 

At this point it looks as 
if we have completely trivialized the reduced phase space quantization 
problem of our constrained Hamiltonian system because there is no 
Hamiltonian to be considered and so it seems that we can just choose any 
of the standard kinematical representations for quantizing the phase space
coordinatized by the $q^a,p_a$ and simply use it for $Q^a,P_a$ because 
the respective Poisson algebras are (weakly) isomorphic. However, this is 
not the case. In addition 
to satisfying the canonical commutation relations we want that the multi 
parameter group of automorphisms $\alpha^\tau$ on $\cal D$ be 
represented unitarily on $\cal H$ (or at least a suitable, preferred
one parameter group thereof). In other words, we want that there exists a 
multi parameter group of unitary operators $U(\tau)$ on $\cal H$ such that
$\pi(\alpha^\tau(Q^a))=U(\tau)\pi(Q^a)U(\tau)^{-1})$ and similarly for 
$P_a$.

Notice that due to the 
relation (which is exact on the constraint surface)
\be \label{3.4}
\alpha^\tau(Q^a)=F_{\alpha'_\tau(q^a),T}
=\sum_{\{k\}} \;\prod_j\; \frac{\tau_j^{k_j}}{k_j !}
F_{\prod_j X_j^{k_j}\cdot q^a,T}
\ee
and where on the right hand side we may replace any occurence of $P_j, 
T_j$ by functions of $Q^a,P_a$ according to the above rules. Hence the 
automorphism $\alpha^\tau$ preserves the algebra of functions of the 
$Q^a,P_a$, although it is a very complicated map in general and in quantum 
theory will suffer from ordering ambiguities. On the other hand, for short
time periods (\ref{3.4}) gives rise to a quickly converging perturbative 
expansion. Hence we see that the representation problem of $\cal D$ will 
be severly constrained by our additional requirement to implement the 
multi time evolution unitarily, if at all possible. Whether or not this is 
feasible will strongly depend on the choice of the $T_j$.

A possible way to implement the multi -- fingered time evolution 
unitarily is by quantizing the Hamiltonians $H_j$ that generate the 
Hamiltonian flows $\tau_j\mapsto \alpha^\tau$ where $\tau_k=\delta_{jk} 
\tau_j$.
This can be done as follows: The original constraints $C_j$ can be solved 
for the momenta $P^j$ conjugate to $T_j$ and we get equivalent constraints
$\tilde{C}_j=P^j+E_j(q^a,p_a,T_k)$. These constraints have a strongly  
Abelean constraint algebra\footnote{Proof: We must have 
$\{\tilde{C}_j,\tilde{C}_k\}=\tilde{f}_{jk}\;^l \tilde{C}_l$ for some
new structure functions $\tilde{f}$ by the first class property. The left 
hand side is independent of the functions $P^j$, thus must be the right 
hand side, which may therefore be evaluated at any value of $P^j$. Set
$P^j=-E_j$. $\Box$.}. We may write $C'_j=K_{jk} \tilde{C}_k$ for some 
regular matrix $K$. Since $\{C'_j,T_k\}\approx \delta_{jk}=
\{\tilde{C}_j,T_k\}$ it follows that $K_{jk}\approx \delta_{jk}$.
In other words $C'_j=\tilde{C}_j+O(C^2)$ where 
the notation $O(C^2)$ means that the two constraints set differ
by terms quadratic in the constraints. It follows that the Hamiltonian
vector fields $X_j,\;\tilde{X}_j$ of $C'_j,\tilde{C}_j$ are weakly 
commuting. We now set $H_j(Q^a,P_a):=F^0_{E_j,T}\approx 
E_j(F^0_{q^a,T},F^0_{p_a,T},F^0_{T_k,T})\approx E_j(Q^a,P_a,0)$. 
Let now $f$ be any function which depends only $q^a,p_a$. Then we have 
\ba \label{3.5}
\{H_j,F^0_{f,T}\} &\approx& 
F^0_{\{E_j,f\}^\ast,T} =F^0_{\{E_j,f\},T} 
=F^0_{\{\tilde{C}_j,f\},T} 
\nonumber\\
&=& 
\sum_{\{k\}} \prod_l \frac{(\tau_l-T_l)^{k_l}}{k_l !} 
\prod_l X_l^{k_l} \cdot \tilde{X}_j \cdot f
\nonumber\\
&\approx & 
\sum_{\{k\}} \prod_l \frac{(\tau_l-T_l)^{k_l}}{k_l !} 
\tilde{X}_j \cdot \prod_l X_l^{k_l} \cdot  f
\nonumber\\
&\approx & \tilde{X}_j\cdot F^0_{f,T}
- \sum_{\{k\}} 
(\tilde{X}_j \cdot \prod_l \frac{(\tau_l-T_l)^{k_l}}{k_l !} )
 \prod_l X_l^{k_l} \cdot  f
\nonumber\\
&\approx & 
+\sum_{\{k\}} \prod_l \frac{(\tau_l-T_l)^{k_l}}{k_l !} \;
 X_j \cdot \prod_l X_l^{k_l} \cdot  f
\nonumber\\
&=&
(\frac{\partial}{\partial \tau_j})_{\tau=0} \alpha^\tau(F_T(f))
\ea
where we have used in the second step that $\{T_j,E_k\}=\{T_j,f\}=0$,
in the third we have used that $\{P_j,f\}=0$, in the fifth we have 
used that the $X_j,\tilde{X}_k$ are weakly commuting, in the seventh we 
have used that $F^0_{f,T}$ is a weak observable, and in the last the 
definition of the flow. We conclude that the Dirac observables $H_j$ 
generate the multifingered flow on the space of functions of the 
$Q^a,P_a$ when restricted to the constraint surface. The algebra of the 
$H_j$ is weakly Abelean because the flow $\alpha^\tau$ is a weakly 
Abelean group of automorphisms. 

Thus, the problem of implementing the flow unitarily can be reduced to 
finding a self adjoint quantization of the functions $H_j$. Preferred 
one parameter subgroups will be those for which the corresponding 
Hamiltonian generator is bounded from below. Notice, however, that in 
(\ref{3.5})
we have computed the infinitesimal flow at $\tau=0$ only. For an arbitrary 
value of $\tau$ the infintesimal generator $H_j(Q^a,P_a,\tau)$ defined by 
\be \label{3.6}
\{H_j(\tau),F^\tau_{f,T}\}:=
\frac{\partial}{\partial \tau_j} \alpha^\tau(F_T(f))
\ee
may not coincide with $F^0_{E_j,T}$ since the Hamiltonian could be 
explicitly time $\tau$ dependent. In particular, the calculation 
(\ref{3.5}) does not obviously hold any more even by setting 
$H_j(\tau):=F^\tau_{E_j,T}$ because even if $f$ depends on $q^a,p_a$ only,
$\alpha'_\tau(f),\alpha'_\tau(E_j)$ may depend on $P_j$ as well.

\section{Reduced Phase Space Quantization of Geometry and Matter}
\label{s4}

In what follows we sketch a possible application of these ideas
to field theory coupled to gravity. Details will appear elsewhere.\
\\
One would like to apply this formalism to the theory which presently 
describes most accurately what we observe, namely General Relativity
for geometry coupled to the Standard Model for matter. In its 
canonical formulation coupled to gravity, see e.g. \cite{4}, we encounter
the following set of constraints: The electroweak $U(1)\times 
SU(2)_L$
Gauss constraints (before symmetry breaking), the QCD $SU(3)$ Gauss 
constraint, the $SU(2)$ Gauss constraint for geometry, the spatial 
diffeomorphism constraint and the Hamiltonian constraint. The various 
Gauss constraints are rather easy to cope with and we will focus on the 
latter four constraints which are, roughly speaking, the Hamiltonian 
incarnation of the 
generators of the four dimensional diffeomorphism group under which
the theory is invariant. The point of view is that we carry out a reduced 
phase space quantization with respect to these constraints and solve the 
remaining Gauss constraints after that. This is possible because the 
Gauss constraints Poisson commute with the four other constraints.

The most natural choice for clock variables are scalar fields and 
remarkably there are precisely four real scalars in the standard model:
The real and imaginary part of the two components of the Higgs dublett.
(Notice however that the Higgs field still awaits its observation).
The Higgs field consists of two complex scalar fields $\phi_I,\;I=1,2$ 
and given a complete orthonormal basis $b_j$ of $L_2(\sigma,d^3x)$ where 
the spacetime manifold $M$ is assumed to be diffeomorphic to $\Rl\times 
\sigma$, we can form the 4 functions 
%(alternatively we use the conjugate momenta)
\be \label{4.1}
T^{1I}_j=\Re(<b_j,\phi_I>),\;T^{2I}_j=\Im(<b_j,\phi_I>)
\ee
where the inner product denoted is that on $L_2(\sigma,d^3x)$.
The four sets of constraints $C_{\mu j}:=<b_j,C_\mu>$ where  
$C\mu,\mu=1,2,3$ stands for the spatial diffeomorphism constraint 
and $C_0=C$ for the Hamiltonian constrainet for the combined matter and 
geometry system, are algebraic expressions in the momenta conjugate to the 
$T^{\alpha I}_j$ and thus can be solved in terms of them. 
Interestingly, for the appropriate choice of sign, the corresponding 
Hamiltonians $E_j$ are {\it positive} functions (which however does not 
imply that the corresponding $H_j(\tau)$ remains positive for all $\tau$). 
Moreover,
the Higgs field and its conjugate momentum have vanishing Poisson brackets 
with the remaining gauge fields (including gravity) and the fermions
(leptons and quarks). 

This is precisely the situation pictured in the 
previous section and we can now start applying the formalism. In 
particular, we would need to look for representations of these remaining 
fields which have a unitary implementation of the ``Higgs Time 
evolution''. Notice that we could use the standard kinematical 
representations for these fields while having already accounted for the 
Hamiltonian constraint which is very difficult to solve in the Dirac
procedure of solving the constraints at the quantum level. The Gauss
constraints mentioned above can be solved within these representations
because the coummute with the constraints $C_{\mu j}$.

Notice that in this picture the Higgs field drops out 
from quantization which looks bad, see the discussion in the next 
section\footnote{One might speculate that  
this could explain why it has not been directly observed yet, however,
it is hard to imagine how the already successful calculations within the 
electroweak theory and which use the Higgs interactions in an essential 
way could be accounted for otherwise.}. 
The clock times $\tau_j$ to be inserted in the 
formalism are the measured values of what one usually calls the Higgs 
expectation value and that one 
uses in the spontaneous symmetry breaking picture and which find its value 
into the masses and couplings of the standard model. Notice also that 
for every function on phase space $f$ which is gauge invariant under the 
electroweak gauge group fails to do so after we apply the map $F_T$ which  
effectively replaces the dependence of the Higgs field in $f$  
by its nondynamical ``expectation value'' $\tau_j$. Thus, the map $F_T$ 
also accomplishes for the spontaneous symmetry breaking. In order that 
the Higgs field serves as a good clock not only must the corresponding 
matrix
$A_{jk}$ be invertible (which is actually the case, it is proportional
to its corresponding conjugate momentum which we assume to be non 
vanishing) but also that the Higgs field values $\tau_j$ should really 
evolve in 
nature. Thus, one expects that masses of leptons and quarks are evolving,
if only very slowly, if the Higgs field is to be a good clock.

\section{Dirac Quantization of Dirac Observables}
\label{s5}

Reduced Phase Space quantization is not what one usually does, mostly one 
quantizes before reducing. This is done because the usual belief is
that the algebra of Dirac observables, if they can be found at all 
classically, is too complicated in order that one can control its 
representation theory. Therfore one starts with a 
kinematical representation of the full unconstrained phase space which 
supports the constraints as (densely defined and closable) operators,
determines the joint (generalized) kernel of the quantum constraints,
computes the induced inner product on that kernel which then becomes 
the physical Hilbert space and finally represents 
the (weak) Dirac observables as self adjoint operators on the physical
Hilbert space. Here weak Dirac observables are operators which preserve,
in an appropriate (generalized) sense, the kernel of the quantum 
constraints. 

On the other hand, we have seen in section \ref{s3} that the machinery of 
\cite{1} allows one to find easily representations of at least a subclass
of Dirac observables if the convergence issues mentioned can be resolved.  
In particular, in the reduced phase space quantization of the algebra of 
the $Q^a,P_a$ sketched in section \ref{s3}, the representation that one 
chooses is already the physical Hilbert space and the Dirac observables 
are represented by self -- adjoint operators there. Thus it seems that 
the reduced phase space quantization is preferred as it circumvents to
work with the representation problem of unphysical quantities altogether.
One could object that while the construction of \cite{1} works in 
principle it is rather involved and $F^\tau_{f,T}$ can hardly be computed 
explicitly. However, as we must deal with the $F^\tau_{f,T}$ also in the 
Dirac quantization picture in order to arrive at physical predictions,
we are confronted with the same problem also in the constraint 
quantization programme. 

There is, however, a difference between the 
constraint quantization programme and the reduced phase space quantization 
programme which leads to physically different predictions: 
As we have seen, in the reduced phase space programme the clock variables
$T_j$ are replaced by real numbers $\tau_j$ and their conjugate momenta 
by the functions $E_j(q^a,p_a,\tau_j)$. Thus $T_j$ is not quantized.
On the other hand, in 
the Dirac quantization programme also the clock variables $T_j$ are 
quantized as well as the conjugate momenta $P^j$ which are not replaced, 
via the constraints, in terms of $q^a,p_a,T_j$. Hence, the representations 
of the Dirac observables that come from constraint quantization know about 
the quantum fluctuations of $T_j$ while those of the reduced phase 
space quantization do not. In particular, different choices of clocks 
will lead to representations which suppress fluctuations of different
clocks. In that sense, the constraint quantization is universal 
because it treats all variables on the same (quantum) footing. For 
instance, using the Higgs field as a clock along the lines of section 
\ref{s4} and using a constraint quantization procedure would allow the 
Higgs field to fluctuate which should be the case as one uses Feynman
diagrammes involving Higgs vertices quite successfully in order to compute 
electroweak processes that are actually measured at CERN. We conclude that
the reduced phase space quantization of the Dirac observables can be 
useful only in a regime where the clocks $T_j$ can be assumed to behave
classically. This is of course not the case with respect to any choice of 
clocks in extreme situations that we would like to access in quantum 
gravity such as at the big bang. There we necessarily need a constraint 
quantization of the system. 

Notice, however, that the 
additional representation problem of implementing the multi -- fingered 
time evolution unitarily in the reduced phase space picture is also
present in the Dirac quantization picture when we try
to simply find an ordering of the quantities $Q^a,P_a$ and 
$\alpha^t(Q^a),\;\alpha^t(P_a)$. This problem can be resolved in the same 
manner, namely by trying to find a self adjoint quantization of the 
Dirac observables corresponding to the Hamiltonians $H_j$ defined in 
section \ref{s3}.

\section{Reducing in Steps and Master Constraint}
\label{s6}

In \cite{1} it was shown that, remarkably, if one is given a constraint 
algebra of the form\footnote{In \cite{1} the specific form of (\ref{6.1})
was not assumed, it is enough that the $C_I$ form a subalgebra. However,
here we are interested in the application to GR only.} 
\be \label{6.1}
\{C_J,C_K\}=f_{JK}\;^L C_L,\;\;
\{C_J,C_k\}=f_{Jk}\;^l C_l,\;\;
\{C_j,C_k\}=f_{jk}\;^L C_L
\ee
one can construct weak Dirac observables of the form $F^\tau_{f,T}$ 
%using
%the new constraints 
%$C'_j=\sum_k (A^{-1})_{jk} C_k+\sum__K (A^{-1})_{jK} C'_K$  
if one has functions $f,T_j$ which are strong Dirac 
observables with respect to the $C_I$ only. 
%Here $A_{ab}=\{C_a,T_b\}$ with $a=j,J$ and $b=k,K$. 
The obvious application is General Relativity
where the $C_I$ play the role of the spatial diffeomorphism constraints 
while the $C_j$ play the role of the Hamiltonian constraint. The former
close with structure constants so that it makes sense to compute 
strong Dirac observables rather than weak ones with respect to the spatial 
Diffeomorphism constraint.

So the statement is that we need only to choose clock variables 
$T_j$ with respect to the second set of constraints $C_j$ if $f,T_j$
are already invariant with respect to the first set of the $C_I$. This is
quite remarkable because the $X_{j_1}..X_{j_n} f$ are not obviously 
invariant under the $C_I$. 
%The proof rests again on the fact that the 
%Hamiltonian vector fields of the $C_I$ and $C'_j=\sum_k (A^{-1})_{jk} C_k$
%are weakly Abelean when evaluated on $C_I-$invariant functions.

In this section we report on a related observation which results from 
the considerations in \cite{5} which is to overcome the dificult problem 
of quantizing constraint algebras of the form (\ref{6.1}) with structure 
functions $f_{jk}\;^l$ rather than structure functions and such that the 
$C_I$ do not form an ideal. The idea
of \cite{4} is to construct the (partial) Master constraint
\be \label{6.2}
M:=\frac{1}{2}\sum_{j,k} Q_{jk} C_j C_k
\ee
where $Q_{jk}$ is a positive, symmetric matrix valued function on phase 
space. The Master Constraint $M=0$ imposes all the individual constraints
$C_j=0$ simultaneously and the condition $\{F,\{F,M\}\}_{M=0}$
is equivalent to the condition that $\{F,C_j\}\approx 0\;\forall j$ is 
a weak Dirac observable with respect to the $C_j$. In application to 
General Relativity it is important that one uses a matrix $Q_{jk}$ such 
that the Master Constraint is invariant with respect to the $C_I$ so that
one can first solve the $C_I$ constraints in quantum theory and then 
the Master constraint (the Master constraint should leave the kernel with 
respect to the $C_I$, that is, the associated spatially diffeomorphism 
invariant Hilbert space, invariant). 

It turns out that the Master constraint is very useful in the quantum 
theory \cite{6} but so far in the classical theory it has not yet been 
possible to construct weak Dirac observables directly using the 
condition $\{F,\{F,M\}\}_{M=0}$. One might think that one should 
simply construct $C':=M/\{M,T\}$ for a single Master clock function 
$T$ and then apply the machinery of \cite{1} however, the corresponding
functions are ill -- defined on the constraint surface in general.
Thus one must use a different method for instance the one developed in 
\cite{1}.

However, one can fruitfully combine the methods of \cite{1} and \cite{5}
as follows: Consider $C_I-$invariant functions 
$T_j$ and a $C_I-$invariant Master constraint $M$. We may now define 
new constraints
\be \label{6.3}
\tilde{C}_j:=\{M,T_j\}\approx \sum_{kl} Q_{kl} C_k A_{lj}
\ee
which are equivalent to the old ones since the matrices $Q,A$ are 
invertible by assumption. Interestingly, the new constraints are 
$C_I-$invariant, $\{C_I,\tilde{C}_j\}=0$ and the constraint algebra 
(\ref{6.1}) can be simplified to
\be \label{6.4}
\{C_J,C_K\}=f_{JK}\;^L C_L,\;\;
\{C_J,\tilde{C}_k\}=0,\;\;
\{\tilde{C}_j,\tilde{C}_k\}=
\tilde{f}_{jk}\;^L C_L+
\tilde{f}_{jk}\;^l C_l
\ee
This means that, using the Master constraint, in {\it General Relativity 
we may 
generate new Hamiltonian constraints which are spatially diffeomorphism 
invariant}. Therefore now the spatial diffeomorphism constraints form an 
ideal and it is possible to first solve the spatial diffeomorphism 
constraints and then to impose the Hamiltonian constraints on the 
spatially diffeomorphism invariant Hilbert space where they then 
close among themselves (with structure functions).
%mustcommute according to (\ref{6.4}). 
In other words,
using the above procedure one can use the philosophy of \cite{5}
to work on the spatially diffeomorphism invariant Hilbert space while 
still using an infinite number of constraints rather than a Master 
constraint. However, unless the $T_j$ and $Q_{jk}$ can be chosen in such a 
way that $\tilde{f}_{jk}\;^l=0$ the algebra still contains structure 
functions in contrast to the Master constraint proposal of \cite{5} which
contains only structure constants which 
might make the quantization of that algebra difficult. 
%Their algebra then simplifies drastically, namely it becomes 
%Abelean.

Nevertheless, at the classical level, using the $\tilde{C}_j$ it is 
possible to choose 
$C_I-$invariant clock 
variables $T_j$ just with respect to the $C_j$ and still 
$F^\tau_{f,T}$ is a weak Dirac observable with respect to {\it 
all} constraints. The proof is simpler than the one in \cite{1}
and goes as follows: Cconsider the yet different but equivalent set of  
constraints
$\tilde{C}'_j=\sum_k (\tilde{A}^{-1})_{jk} \tilde{C}_k$,
where 
$\tilde{A}_{jk}=\{\tilde{C}_j,T_k\}$ is a non - degenerate
(symmetric, if $\{T_j,T_k\}=0$) matrix which is now $C_I-$invariant as 
well.
It is clear that the corresponding functions $F^\tau_{f,T}$ are 
exactly $C_I-$invariant since the $X'_{j_1}..X'_{j_n}\cdot f$ are. We thus 
just have to show that the Hamiltonian vector fields 
$\tilde{X}_j'$ of the 
$\tilde{C}'_j$ are weakly Abelean when applied to $C_I-$invariant 
functions $f$. This follows from a calculation similar to (\ref{2.6}):
Abbreviating $\tilde{B}_{jk}:=(\tilde{A}^{-1})_{jk}$ we now have,
following exactly the same steps   
\ba \label{6.6}
&& \{\tilde{C}'_j,\{\tilde{C}'_k,f\}\}-\{\tilde{C}'_k,\{\tilde{C}'_j,f\}\}
\nonumber\\
%&=&\sum_{m,n} \tilde{B}_{jm}[-\sum_{l,i} \tilde{B}_{kl} \tilde{B}_{in} 
%\{\tilde{C}_n,f\} 
%(\{\tilde{C}_m,\{\tilde{C}_l,T_i\}\}-\{\tilde{C}_l,\{\tilde{C}_m,T_i\}\}) 
%+\tilde{B}_{kn} 
%(\{\tilde{C}_m,\{\tilde{C}_n,f\}\}-\{\tilde{C}_n,\{\tilde{C}_m,f\}\})]
%\nonumber\\
&=&\sum_{m,n} \tilde{B}_{jm}[\sum_{l,i} \tilde{B}_{kl} \tilde{B}_{in} 
\{\tilde{C}_n,f\} 
\{T_i,\{\tilde{C}_m,\tilde{C}_l\}\} -\tilde{B}_{kn} 
(\{f,\{\tilde{C}_m,\tilde{C}_n\}\}]
\nonumber\\
&\approx &\sum_{m,n} \tilde{B}_{jm}[-\sum_{l,i,p} \tilde{B}_{kl} 
\tilde{B}_{in} \{\tilde{C}_n,f\} 
\tilde{f}_{ml}\;^p \tilde{A}_{pi}+\tilde{B}_{kn}\sum_l \tilde{f}_{mn}\;^l 
\{\tilde{C}_l,f\}]
\nonumber\\
&= &\sum_{m,n,l} \tilde{B}_{jm}[-\tilde{B}_{kl} \{\tilde{C}_n,f\} 
\tilde{f}_{ml}\;^n +B_{kn} \tilde{f}_{mn}\;^l \{C_l,f\}]
=0
\ea
where the terms proportional to $\{C_I,T_j\},\;\{C_I,f\}$, which appeared 
at an intermediate stage in the third step, drop out exactly due to the 
invariance of $T_j,f$ under the $C_I$.

\section{Conclusions}
\label{s7}

The proposal of \cite{1} shows that the issue of the construction of 
Dirac observables for General Relativity is not as hopless as it seems.
While there are many open issues even in the classical theory such as 
convergence and differentiability 
of the formal power series constructed, we now have analytical expressions 
available and these can be used in order to make the framework rigorous
in principle. Physical insight will be necessary in order to identify 
the mathematically most convenient and physically most relevant clocks
especially for field theories such as General Relativity. 

In the quantum 
theory, either in the reduced phase space picture or the Dirac constraint 
quantization picture, the challenge will be to construct the corresponding 
self -- 
adjoint operators on the physical Hilbert space as well as the generators
of the multi -- fingered time evolution. This will be very hard but in a
dynamical system as complicated as General Relativity this is to be 
expected. One hopes, of course, that the power series provided in \cite{1}
will help to develop a systematic approximation scheme or perturbation 
theory close to the gauge cut $T=\tau$.\\
\\
In the present paper we have mainly reported some first observations and 
ideas. We hope to fill the many gaps in future publications.

%\newpage

~\\
{\large Acknowledgements}\\
\\
We thank Bianca Dittrich for many helpful explanations, comments and 
discussions.
This research project was supported in part by a grant from NSERC 
of Canada to the Perimeter Institute for Theoretical Physics.

\end{document}